\documentclass[onecolumn,aps,floatfix,pra,superscriptaddress,nofootinbib]{revtex4}

\usepackage{hyperref}
\usepackage{color}
\usepackage{graphicx}
\usepackage{bm}
\usepackage{amsmath}
\usepackage{amsbsy}
\usepackage{enumerate}
\usepackage{upgreek}
\usepackage[subnum]{cases}
\usepackage{dsfont}

\newcommand{\cl}{ \text{cl} }

\newcommand{\tr}{ \text{tr} }
\newcommand{\ti}{ \tilde }
\newcommand{\ep}{ \epsilon }
\newcommand{\pa}{ \partial }
\newcommand{\hb}{ \hbar }
\newcommand{\si}{ \sigma }

\newcommand{\ga}{ \gamma }
\newcommand{\la}{ \langle }
\newcommand{\ra}{ \rangle }

\newcommand{\im}{ \text{Im} }

\newcommand{\free}{ \text{f} }
\newcommand{\nc}{ \text{nc} }
\newcommand{\hbt}{ \tilde{\hbar} }
\newcommand{\tpsi}{ \tilde{\psi} }
\newcommand{\tphi}{ \tilde{\phi} }

\newcommand{\trho}{ \tilde{\rho} }

\newcommand{\tst}{ \tilde{s}_t }

\newcommand{\wti}{ \widetilde }
\newcommand{\thM}{ \ti{\hat{M}} }
\newcommand{\thH}{ \ti{\hat{H}} }

\newcommand{\tiW}{ \ti{W} }

\begin{document}
\title{ Quantum Classical Transition for Mixed States: The Scaled Von Neumann Equation }
\author{S. V. Mousavi}
\email{vmousavi@qom.ac.ir}
\affiliation{Department of Physics, University of Qom, Ghadir Blvd., Qom 371614-6611, Iran}

\author{S. Miret-Art\'es}
\email{s.miret@iff.csic.es}
\affiliation{Instituto de F\'isica Fundamental, Consejo Superior de Investigaciones Cient\'ificas, Serrano 123, 28006 Madrid, Spain}

\begin{abstract}

In this work, we proposed a smooth transition wave equation from a quantum to classical regime in the 
framework of 
von Neumann formalism for ensembles and then obtained an equivalent scaled equation. This led us to 
develop a scaled statistical theory following the well-known Wigner--Moyal approach of quantum 
mechanics. This scaled nonequilibrium statistical mechanics has in it all the ingredients of the classical and 
quantum theory described in terms of a continuous parameter displaying all the dynamical regimes in between 
the two extreme cases. Finally, a simple application of our scaled formalism consisting of  reflection from a mirror by computing 
various quantities, including probability density plots, scaled trajectories, and arrival times, was analyzed.

\end{abstract}

\maketitle

%%%%%%%%%%%%%%%%%%%%%%%%%%%%%%%%%%%%%%%%%%

\section{Introduction}

The most general formulation of quantum mechanics is given in terms of a density operator, which is a statistical mixture 
of state vectors. Furthermore, in an open or composite quantum system, the system of 
interest is described by the {{reduced}} density operator, which is obtained by tracing out the total 
density operator over the remaining degrees of freedom.%MDPI; please confirm if italic should remain, same for highlights below.
%
%The meaning of density matrix was the subject of some studies~\cite{AnAh-FPL-1999}. 
Using the method of protective measurements, Anandan and Aharonov have proposed the observation of 
the density matrix of a {{single}} system, thus presenting a new meaning of the density matrix in this context~\cite{AnAh-FPL-1999}.
In~this regard, it has been shown that the density matrix can be consistently treated as a property of an individual system, not of an ensemble alone~\cite{Ma-FP-2005}.
In~addition to the statistical (mixture) and reduced density matrices, the conditional density matrix, 
which is conditional on the configuration of the environment, has been discussed~\cite{DuGoTuZa-FP-2005} and  
argued that the precise definition is possible only in \mbox{Bohmian mechanics}.

Standard quantum mechanics is unable to provide an explanation for the non-appearance of macroscopically distinguishable states. Different approaches have been adopted in the literature to overcome this problem, the measurement problem, the reduction or the collapse postulate, and the localization of the wave packet. 
Environmental decoherence theories~\cite{JoZeKiGiKuSt-book-2002, Sch-book-2007, Zu-PT-1991} seek 
an explanation entirely within the standard quantum mechanics while taking into account the crucial role played 
by the environment of the quantum system. The whole system evolves under the usual  Schrödinger 
equation. Then, by tracing over the environmental degrees of freedom, a master equation is obtained for 
the reduced density operator describing the system of interest, which contains parameters such as the 
friction coefficient and the temperature of the environment. 
In the second approach to the problem, the Schrödinger equation is modified in such a way that coherence 
is automatically destroyed when approaching to the macroscopic level. This has been called 
``intrinsic`` decoherence by Milburn~\cite{Mi-PRA-1991}. Perhaps the model introduced by  Ghirardi, 
Rimini, and Weber~\cite{GRW-PRD-1986} is the most widely known in this connection, where two 
fundamentally distinct evolution equations of the standard quantum mechanics, i.e., the unitary time 
evolution given by the Schrödinger equation in the absence of measurements and the irreversible 
collapse rule, apply during a measurement and are merged in a unique \mbox{dynamical description}. %Please check meaning retained

On the other hand, Bohmian mechanics~\cite{Holland-book-1993, salva1, salva2, NaMi-book-2017} are clearly 
a complementary, alternative, and new interpretive way of introducing quantum mechanics, {wherein they provide a clear 
picture of quantum phenomena in terms of trajectories in configuration space}. A~ smooth 
transition could then be devised {by considering the quantum classical transition differential 
wave equation, due to Richardson et al}.~\cite{RiSchMaVaBa-PRA-2014} for 
conservative systems. %MDPI: There are some overlaps with the existing work, please make careful revisions. The following highlights are the same.
In doing~so, the corresponding dynamics are governed by a continuous parameter, the  transition parameter, 
which leads to a continuous description of any quantum phenomena in terms of trajectories and  scaled trajectories
\cite{MoMi-AP-2018,MoMi-JPC-2018}. In ~other words, one is, thus, able to describe any dynamical regime in between
the quantum and classical ones in a continuous way by emphasizing how this smooth process is established
({a scaled Planck's constant can then be  defined from the transition parameter covering the lim}it $ \hbt \to 0$).
{Doing so, scaled trajectories also display the well-known non-crossing property even in the classical regime}. 
Chou  applied this wave equation to analyze wave--packet interference~\cite{Chou1-2016} and the dynamics of 
the harmonic and Morse oscillators with complex trajectories~\cite{Chou2-2016}. Stochastic Bohmian
and scaled trajectories have also been discussed in the literature for open quantum systems~\cite{MoMi-FOP-2022}.
{Moreover, if  a time-dependent Gaussian ansatz is assumed for the probability density,
Bohmian and scaled trajectories are expressed as a sum of  a classical trajectory (that is, a particle feature) 
and a term containing the width of the corresponding wave packet (that is, a wave feature), which has been called the} {{dressing scheme}}~\cite{NaMi-book-2017}. Analogously, this  scheme 
has also been observed in the context of nonlinear quantum mechanics~\cite{Feng}, where, for example, solitons 
also {display this wave--particle duality; appearing their wave property in the form of a travelling solitary 
wave, and their corpuscle feature is analogous to a classical partic}le. %Interesting enough,

%(in this regime, initial momenta are 
%determined through a guidance equation from initial positions).
%

{The procedure of using a continuous parameter to smoothly monitor the different dynamical regimes in the theory 
recall us  the well-known WKB approximation, widely used for conservative systems. 
Several important differences are worth stressing. First, the classical Hamilton--Jacobi equation for the 
action is obtained at zero order in the $\hbar$-expansion, whereas the so-called classical wave (nonlinear) }
equation~\cite{Schi-PR-1962} is reached by construction. {Second, the hierarchy of the differential equations 
for the action at different orders of the same expansion}  is substituted by only a transition differential 
{wave equation, which can be  solved in the linear and nonlinear domains. 
Third, the quantum to classical transition for trajectories is carried out in a continuous and gradual 
way, thus stressing the different dynamical r}egimes in between the quantum and classical ones. 
Fourth, this continuous and smooth transition could also be thought as a gradual decoherence process (let us say, 
internal decoherence; in this context, the decoherence process is used only to stress the fact that we 
are approaching the classical limit, and no measurement is carried out) due to the scaled Planck constant, 
thus allowing us to analyze what happens at intermediate regimes~\cite{MoMi-JPA-2023}.
However, decoherence effects have also been analyzed in  interference phenomena using 
a class of quantum trajectories based on the same grounds as Bohmian ones, which is associated with the 
system-reduced density matrix~\cite{SaBo-EPJD-2007}. Such a study has been carried out for statistical 
mixtures and studied in the framework of Bohmian mechanics~\cite{LuSa-AP-2015} regarding the minimal view, 
i.e., without any reference to the quantum potential.
Even more, by writing the density matrix in the polar form, a Bohmian trajectory formulation for 
dissipative systems has been proposed, where a {{double quantum potential}} being a measure of the 
local curvature of the density amplitude is responsible for quantum effects~\cite{MaBi-JCP-2001}.
A different approach has been taken for the hydrodynamical formulation of mixed 
states~\cite{BuCe-JCP-2001}, where a local-in-space formulation has been adopted in the sense that a 
hierarchy of moments contains the non-local information associated with the off-diagonal elements of the 
density matrix.

In the present article, our purpose is to provide a clear formulation of the pure and mixed ensembles in terms 
of Bohmian mechanics by using the polar form of the density matrix within the von Neumann equation
framework. In ~this way, the corresponding quantum potential is introduced, and a momentum vector field is 
defined for both forward and backward in time motions. Once this is carried out,
within  the quantum classical transition equation framework, a scaled  Schrödinger equation is easily 
derived that leads to the so-called scaled von Neumann equation. Afterwards, Moyal's procedure~\cite{Mo-PCPS-1949}
used to interpret quantum mechanics as a statistical theory is then applied to the scaled theory by considering 
a characteristic function, which is a standard function in statistical mechanics~\cite{Hi-arXiv-2014}.
The expectation value of the so-called Heisenberg--Weyl operator~\cite{Hi-JCE-2015} is treated as 
the characteristic function. Then, the inverse Fourier transform of the characteristic function is considered 
as the probability distribution function, and its time evolution is thus obtained. % the standard methods of %statistical mechanics. 
In this way, the classical Liouville equation is again derived within this scaled theory. 
The foundations of nonequilibrium statistical mechanics are based on the Liouville equation, 
which is associated with Hamiltonian dynamics (in general, in phase space). In~other words, with this theoretical analysis, we have clearly shown that scaled statistical mechanics are well
established and ready to be applied. This new nonequilibrium statistical mechanics would be valid for any 
dynamical regime, going from the quantum to the classical ones. 
As a simple illustration of our new formulation, scattering a statistical mixture of Gaussian wave packets 
from a hard wall, was studied and compared to the corresponding superposed state. The application of the 
new scaled nonequilibrium statistical mechanics will be postponed for use in future work.

\section{Theory}

When an isolated physical system is described by a density operator $\hat{\rho}$ instead of a state vector 
$ | \psi \rangle$, the equation of motion ruled by the density operator is the so-called Liouville--von Neumann 
equation or simply the von Neumann equation, which is \mbox{written as}
\begin{eqnarray} \label{eq: Neumann}
	i \hb \frac{\pa \hat{\rho}}{\pa t} &=& [\hat{H}, \hat{\rho}]  ,
\end{eqnarray}
where $ \hat{H} $ is the Hamiltonian operator of the system, and $[\cdot, \cdot]$ represents the commutator of two operators. 
For a single particle and in one dimension, this Hamiltonian is expressed as 
\begin{eqnarray}
	\hat{H} &=& \frac{ \hat{p}^2 }{2m} + U( \hat{x} )   ,
\end{eqnarray}
where the first term is the kinetic energy operator, and the second one is the external interaction potential, $U( \hat{x} )$.
This equation reads as
%
%\section{Bohmian formalism for ensembles} \label{sec: vN}
%
%Mixed as well as pure ensembles are described by density operator whose evolution is governed by the von Neumann equation
%
%\begin{eqnarray} \label{eq: vn-abs}
%i\hb \frac{\pa \hat{\rho}}{\pa t} &=& [\hat{H}, \hat{\rho}] , \qquad 
% \hat{H} = \frac{ \hat{p}^2 }{2m} + U(\hat{x})
%\end{eqnarray}
%
%which reads 
%
\begin{eqnarray} \label{eq: vn_eq}
i\hb \frac{\pa}{\pa t} \rho(x, y, t) &=& - \frac{\hb^2}{2m} 
\left( \frac{\pa^2}{ \pa x^2 } - \frac{\pa^2}{ \pa y^2 }   \right) \rho(x, y, t) 
+ (U(x) - U(y)) \rho(x, y, t)   
\end{eqnarray}
in the position representation. Diagonal elements of the density matrix give probabilities, while the non-diagonal elements represent coherences. 

\subsection{Pure Ensembles in the de Broglie--Bohm Approach}

Before considering mixed ensembles, it is important first to look at pure ones in the framework of the von 
Neumann equation. Density operator elements, in coordinate representation, for the pure state 
$ \hat{\rho} = | \psi \ra \la \psi | $ are given by $ \rho(x, y, t) = \psi(x, t) \psi^*(y, t), $ where the wave function 
$ \psi(x, t) $ is governed by the Schrödinger equation
\begin{eqnarray} \label{eq: Sch_eq}
	i\hb \frac{\pa}{\pa t} \psi(x, t) &=& \left( - \frac{\hb^2}{2m}  \frac{\pa^2}{ \pa x^2 } + U(x) \right) \psi(x, t)  ,
\end{eqnarray}
and its complex conjugate $ \psi^*(y, t) $ is governed by the complex conjugation of the same equation, i.e., 
\begin{eqnarray} \label{eq: Sch_eq-cc}
	- i\hb \frac{\pa}{\pa t} \psi^*(y, t) &=& \left( - \frac{\hb^2}{2m}  \frac{\pa^2}{ \pa y^2 } + U(y) \right) \psi^*(y, t)  .
\end{eqnarray}

%MDPI: new indent, please confirm, same for highlights below.
{This }equation reveals that $ \psi^*(y, t) $ is the wave function corresponding to the {time-reversed} 
state~\cite{Sakurai-book-1994}. By writing the wave function in its polar form
\begin{eqnarray} \label{eq: psi_polar}
	\psi(x, t) &=& a(x, t) ~ e^{i s(x, t) / \hb}   ,
\end{eqnarray}
$ a(x, t) $ and $ s(x, t) $ are both real functions of the amplitude and phase of the wave function, respectively. 
By substituting this polar form in Equation \eqref{eq: Sch_eq} and splitting the resultant equation in its real and 
imaginary parts, one {obtains}
\begin{numcases} ~
	-\frac{\pa}{\pa t} s(x, t) = \frac{1}{2m} \left( \frac{\pa}{\pa x} s(x, t) \right)^2 + U(x) + q(x, t)
	\hspace{11pt}\qquad\qquad\qquad\qquad \label{eq: HJ-psi}
	\\ 
	\frac{\pa}{\pa t} (a(x, t))^2 + \frac{\pa}{\pa x} \left( (a(x, t))^2 \frac{1}{m} \frac{\pa}{\pa x} s(x, t) \right) = 0 , \label{eq: con-psi}
\end{numcases}
which are respectively the generalized Hamilton--Jacobi and the continuity equations, where 
\begin{eqnarray} \label{eq: qp-psi}
	q(x, t) &=& - \frac{\hb^2}{2m} \frac{1}{a(x, t)} \frac{\pa^2}{\pa x^2} a(x, t)
\end{eqnarray}
is the well-known {quantum potential}. These equations suggest the definition of the momentum field as
\begin{eqnarray} \label{eq: mom-field_x}
	p(x, t) &=& \frac{\pa}{\pa x}s(x, t) 
\end{eqnarray} 
and the velocity field as 
\begin{eqnarray} \label{eq: vel-field_x}
	v(x, t) &=& \frac{1}{m} p(x,t)   .
\end{eqnarray} 
%
%$ v(x, t) = p(x, t) / m $.

{Bohmian} trajectories $ x(x^{(0)}, t) $ are, thus, constructed from the {guidance equation} as
\begin{eqnarray} \label{eq: traj-pure_x}
	\frac{d x}{dt} &=& v(x, t) \bigg|_{x = x(x^{(0)}, t)}   ,
\end{eqnarray} 
with $ x^{(0)} $ being the initial position.

By applying the operator $ \pa_x $ to Equation \eqref{eq: HJ-psi} and using Equation \eqref{eq: vel-field_x}, one reaches a
Newtonian-like equation according to
\begin{eqnarray} \label{eq: newton-psi}
	\frac{d}{dt} p(x, t) &=& - \frac{\pa}{\pa x} (U(x) + q(x, t))  ,
\end{eqnarray}
which shows that regarding $q(x, t)$ as a potential on the same footing as $U(x)$ is consistent. 

Now, let us consider the evolution of $ \psi^*(y, t) $. As stated above, its evolution is governed by 
Equation \eqref{eq: Sch_eq-cc} and expressed again in the polar form
\begin{eqnarray} \label{eq: psi*_polar}
	\psi^*(y, t) = a'(y, t) e^{i s'(y, t)/\hb }   .
\end{eqnarray}
{One has that }
\begin{numcases}~
	- \left(- \frac{\pa}{\pa t} \right) s'(y, t) = \frac{1}{2m} \left( \frac{\pa}{\pa y} (s'(y, t)) \right)^2 + U(y) + q'(y, t)
	\hspace{9.5pt}\qquad\qquad\label{eq: HJ-psi*}
	\\
	- \frac{\pa}{\pa t} (a'(y, t))^2 + \frac{\pa}{\pa y} \left( (a'(y, t))^2 \frac{1}{m} \frac{\pa}{\pa y} s'(y, t) \right) = 0  ,
	\label{eq: con-psi*}
\end{numcases}
where
\begin{eqnarray} \label{eq: qp-psi*}
	q'(y, t) &=& - \frac{\hb^2}{2m} \frac{1}{a'(y, t)} \frac{\pa^2}{\pa y^2} a'(y, t)  .
\end{eqnarray}

{By} comparison to Equations \eqref{eq: HJ-psi} and \eqref{eq: con-psi}, $\pa_t$ has been replaced by $-\pa_t$ 
as the result of the invariance or symmetry under time reversal. 
With respect to these new equations, one defines the momentum field, in the $y$ direction, as
\begin{eqnarray} \label{eq: mom-field_y}
	p'(y, t) &=& - \frac{\pa}{\pa y}s'(y, t) ,
\end{eqnarray} 
which, using Equations \eqref{eq: HJ-psi*} and \eqref{eq: con-psi*}, again yields a Newtonian-like equation
\begin{eqnarray} \label{eq: newton-psi*}
	\frac{d}{dt} p'(y, t) &=& \frac{\pa}{\pa y} (U(y) + q'(y, t))  .
\end{eqnarray}

{Note} that the minus sign in Equation \eqref{eq: mom-field_y} reflects the time-reversed dynamics in the $y$ 
direction~\cite{MaBi-JCP-2001}. However, note that a comparison between Equation \eqref{eq: psi_polar} and 
Equation \eqref{eq: psi*_polar}{reveals that}
\begin{numcases}~
	a'(y, t) = a(y, t)  \hspace{20pt}\qquad\qquad\qquad\qquad\qquad\qquad\qquad \label{eq: a'a} \\
	s'(y, t) = - s(y, t), \hspace{20pt}\qquad\qquad\qquad\qquad\qquad\qquad\qquad \label{eq: s's}
\end{numcases}
and, as one expects, $ p'(y, t) = - p(y, t) $.

By subtracting Equation \eqref{eq: HJ-psi*} from Equation \eqref{eq: HJ-psi} and using Equations \eqref{eq: a'a} and \eqref{eq: s's},
we have that
\begin{eqnarray} \label{eq: HJ-pure}
	-\frac{\pa}{\pa t} [s(x, t) - s(y, t)] &=&
	\frac{ \{\pa_x [s(x, t) - s(y, t)]\}^2 }{2m} - \frac{ \{\pa_y [s(x, t) - s(y, t)]\}^2 }{2m}  + [ U(x) - U(y) ] 
	\nonumber \\
	& - & \frac{\hb^2}{2m} \frac{1}{a(x, t) a(y, t)} ( \pa_x^2 - \pa_y^2 ) [a(x, t) a(y, t)]   .
\end{eqnarray}
{Multiplying} Equation \eqref{eq: con-psi} by $ (a'(y, t))^2 $ and Equation \eqref{eq: con-psi*} by $ (a(x, t))^2 $ and then subtracting 
the resulting equations and using Equations \eqref{eq: a'a} and \eqref{eq: s's} yield
\begin{eqnarray} \label{eq: con-pure}
	\frac{\pa}{\pa t} [a(x, t)a(y, t)]^2 &+& \frac{\pa}{\pa x} \left( [a(x, t)a(y, t)]^2 \frac{1}{m} \frac{\pa}{\pa x} [s(x, t) - s(y, t)] \right) \nonumber \\
	&-& \frac{\pa}{\pa y} \left( [a(x, t)a(y, t)]^2 \frac{1}{m} \frac{\pa}{\pa y} [s(x, t) - s(y, t)] \right) = 0   .
\end{eqnarray}

{Note} that the real functions $ a(x, t)a(y, t) $ and $ s(x, t) - s(y, t) $ appearing in Equations~\eqref{eq: HJ-pure}~and ~\eqref{eq: con-pure} are the amplitude and phase of the {pure} density matrix, respectively, where 
\begin{eqnarray} \label{eq: rho-pure_polar}
	\rho(x, y, t) &=& \psi(x, t) \psi^*(y, t) = 
	a(x, t) a(y, t) e^{i(s(x, t)-s(y,t))/\hb}    .
\end{eqnarray}

{One }could directly reach Equations (\ref{eq: HJ-pure}) and (\ref{eq: con-pure}) by introducing this polar form in 
the von Neumann equation (\ref{eq: vn_eq}).

\subsection{Mixed Ensembles in the de Broglie--Bohm Framework}

Let us now consider a mixed state. The hermiticity of the density operator $ \hat{\rho} $ implies
\begin{eqnarray} \label{eq: rho-Hermit}
	\rho(x, y, t) &=& \rho^*(y, x, t)  .
\end{eqnarray}

{From} this property and the polar form of the density matrix
\begin{eqnarray} \label{eq: rho_pol}
	\rho(x, y, t) &=& A(x, y, t) e^{ i S(x, y, t) / \hb }   ,
\end{eqnarray}
one has that 
\begin{numcases}~ 
	A(x, y, t) = A(y, x, t) \hspace{30pt}\qquad\qquad\qquad\qquad\qquad\qquad\qquad\qquad\qquad \label{eq: A-property} \\
	S(x, y, t) = - S(y, x, t)    , \hspace{30pt}\qquad\qquad\qquad\qquad\qquad\qquad\qquad\qquad\qquad \label{eq: S-property}
\end{numcases}
i.e., the amplitude  (phase) of the density matrix is symmetric (antisymmetric) under the 
$ x \leftrightarrow y $ interchange.
Now, by introducing Equation \eqref{eq: rho_pol} into the von Neumann equation~(\ref{eq: vn_eq}) and splitting 
the resultant equation in real and imaginary parts, one again obtains the Hamilton--Jacobi equation for the phase
\begin{equation} \label{eq: HJ-mixed}
	-\frac{\pa}{\pa t} S(x, y, t) =
	\frac{ [\pa_x S(x, y, t)]^2 }{2m} - \frac{ [\pa_y S(x, y, t)]^2 }{2m}  + [ U(x) - U(y) ]
	+ Q(x, y, t)   
\end{equation}
and the continuity equation
\begin{eqnarray} \label{eq: con-A1}
\frac{\pa}{\pa t} A(x, y, t) + \frac{1}{m} (\pa_x A~ \pa_x S - \pa_y A~ \pa_y S ) 
+ \frac{1}{2m} A (\pa_x^2 S - \pa_y^2 S ) &=& 0   .
\end{eqnarray}
For the amplitude,
\begin{eqnarray} \label{eq: Qp-mixed}
	Q(x, y, t) &=& - \frac{\hb^2}{2m} \frac{1}{A(x, y, t)} ( \pa_x^2 - \pa_y^2 ) A(x, y, t)
\end{eqnarray}
is again the corresponding quantum potential. By defining  the two-component momentum vector field as
\begin{eqnarray} \label{eq: mom_vec}
	\mathbf{P}(x, y, t) &=& ( \pa_x S(x, y, t), - \pa_y S(x, y, t) )  ,
\end{eqnarray}

{Equation} \eqref{eq: con-A1} can be written in its compact form as
\begin{eqnarray} \label{eq: con-A}
	\frac{\pa}{\pa t} A(x, y, t) +  \mathbf{V}(x, y, t) \cdot \bm{\nabla} A(x, y, t) 
	+ \frac{1}{2} A(x, y, t) ~ \bm{\nabla} \cdot \mathbf{V}(x, y, t)  &=& 0   
\end{eqnarray}
or 
\begin{eqnarray} \label{eq: con-A222}
	\frac{\pa}{\pa t} A(x, y, t) +   \bm{\nabla} \cdot [ A(x, y, t)  \mathbf{V}(x, y, t) ]
	- \frac{1}{2} A(x, y, t) ~ \bm{\nabla} \cdot \mathbf{V}(x, y, t)  &=& 0   ,
\end{eqnarray}
where we have introduced the velocity vector field
\begin{eqnarray} \label{eq: vel_vec}
	\mathbf{V}(x, y, t) &=& \frac{1}{m} \mathbf{P}(x, y, t)  .
\end{eqnarray}

{Equation} \eqref{eq: con-A} can, thus, be written as the usual {{continuity equation}}
\begin{eqnarray} \label{eq: con-mixed}
	\frac{\pa}{\pa t} A(x, y, t)^2 + \bm{\nabla} \cdot [ A(x, y, t)^2 \mathbf{V}(x, y, t) ] &=& 0 
\end{eqnarray}
for the conservation of $ A(x, y, t)^2 $ in the two-dimensional space represented by  $x$ and $y$, where
\begin{eqnarray}
	\frac{d}{dt} \int_{-\infty}^{\infty} \int_{-\infty}^{\infty} dx dy~ A(x, y, t)^2 = 0   .
\end{eqnarray}

Using Equations \eqref{eq: mom_vec} and \eqref{eq: HJ-mixed}, one obtains the quantum Newton-like equation as
\begin{eqnarray} 
	\frac{d}{dt} \mathbf{P}(x, y, t) &=& \left( \frac{\pa}{\pa t} + \mathbf{V}(x, y, t) \cdot \bm{\nabla} \right) \mathbf{P}(x, y, t) \nonumber \\
	&=& - \bm{\nabla} ( U \pm Q )  ,
\end{eqnarray}
where the $+$ ($-$) sign inside the parentheses stands for $x$ ($y$) component of the momentum~field. 

%The continuity equation, which is a conservation equation for $ A(x, y, t)^2 $ in the two-dimensional space represented by coordinates $x$ and $y$, can also be expressed according to

%We know that diagonal elements of the density matrix have the interpretation of probability. Thus, we'll try to find a conservation law for this quantity. 
%By taking the derivatives in Equation (\eqref{eq: con-mixed}) and reordering one obtains

If one uses the center of mass and relative coordinates {according to}
\begin{numcases}~
	R = \frac{x+y}{2} \hspace{40pt}\qquad\qquad\qquad\qquad\qquad\qquad\qquad\qquad\qquad \\
	r = x - y   ,
\end{numcases}

{Equation} \eqref{eq: con-A} is rewritten as
\begin{equation} \hspace{50.5pt}\qquad\qquad\qquad\qquad\label{eq: con-A-Rr}
	\frac{\pa}{\pa t} A(R, r, t) + \frac{\pa}{\pa R} \left( A(R, r, t) \frac{1}{m} \frac{\pa}{\pa r} S(R, r, t) \right) + \frac{1}{m} \left( \frac{\pa}{\pa R} S(R, r, t) \right) \left( \frac{\pa}{\pa r} A(R, r, t) \right) = 0  .
\end{equation}

{Note} that this equation can also be directly obtained from the von Neumann equation in the $(R, r)$ coordinates
\begin{equation} \label{eq: vn_eq-Rr}
	\left[ \frac{\pa}{\pa t} + \frac{\hb}{i m} \frac{\pa^2}{ \pa R \pa r } 
	- \frac{ U(R+r/2) - U(R-r/2) }{i \hb} \right]  \rho(R, r, t)  = 0.
\end{equation}

{From}~Equation \eqref{eq: A-property}, it is seen that $ A(R, r, t) $ is an even function of the relative coordinate $r$, and, 
thus, its derivative with respect to $r$ is odd under $ r \rightarrow -r $. This implies that the last term of 
Equation \eqref{eq: con-A-Rr} is zero for $r=0$, i.e., when considering diagonal elements. From~this analysis, one arrives at
\begin{equation} \label{eq: prob-conser}
	\frac{\pa}{\pa t} A(x, t) + \left\{ \frac{\pa}{\pa R} \left( A(R, r, t) \frac{1}{m} \frac{\pa}{\pa r} S(R, r, t) \right) \right\} \bigg |_{R= x, r=0} = 0   
\end{equation}
for the conservation of probability, i.e., diagonal elements of the density matrix 
$ \rho(R=x, r=0, t) = A(R=x, r=0, t) \equiv A(x, t) $, from which the Bohmian velocity field
is obtained as follows
\begin{equation} \label{eq: BM-vel}
	v(x, t) = \frac{1}{m} \frac{\pa}{\pa r} S(R, r, t) \bigg |_{R= x, r=0} .
\end{equation}

{Note} that this velocity field can also be deduced from the probability current density %\cite{•}
\begin{eqnarray}
	j(x, t) &=& \frac{\hb}{m} \im \left\{ \frac{\pa}{\pa x} \rho(x, y, t)\bigg|_{y=x} \right \}
\end{eqnarray} 
through the ratio $ j(x, t) / \rho(x, t) $.

\subsection{The Scaled Von Neumann Equation: A Proposal for Quantum Classical Transition}

In an effort to describe a quantum-to-classical continuous transition, the following {{non-linear}} transition equation 
was proposed to be in the Schr\`{o}dinger framework %Attention: Spelling of Schrodinger different from elsewhere in manuscript 
\begin{eqnarray}  \label{eq: QC-tran}
i \hb \frac{\pa}{\pa t}\psi_{\ep}(x, t) &=& \left[ -\frac{\hb^2}{2m} 
\frac{\pa^2}{\pa x^2} + U(x) + (1-\ep) \frac{\hb^2}
{2m} \frac{\pa_x^2 |\psi_{\ep}(x, t)|}{|\psi_{\ep}(x, t)|} 
\right] \psi_{\ep}(x, t) , 
\end{eqnarray}
which contains the so-called {{transition parameter}} $\ep$ going from one (quantum regime) to zero 
(classical regime) and in between.  By substituting the polar form 
$ R_{\ep} e^{i S_{\ep}/\hb} $ of the wave function in Equation \eqref{eq: QC-tran}, splitting the resultant 
equation into its real and imaginary parts, and introducing the scaled wave function 
$ \tpsi = \psi_{\ep} e^{ i(1/\sqrt{\ep}-1) S_{\ep} / \hb } $, after some straightforward manipulations, 
one arrives at the equivalent scaled {{linear}} equation 
\begin{eqnarray} \label{eq: Scaled Sch}
i \hbt \frac{\pa}{\pa t} \tpsi(x, t) &=& \left[ -\frac{\hbt^2}{2m} \frac{\pa^2}{\pa x^2} + U(x) \right] \tpsi(x, t) ,
\end{eqnarray}
which was shown elsewhere~\cite{RiSchMaVaBa-PRA-2014}, with the so-called {{scaled Planck constant}} being 
\begin{eqnarray} \label{eq: hbt}
\hbt &=& \sqrt{\ep}~ \hb  .
\end{eqnarray}

{This} study has been generalized to dissipative systems in the framework of the Caldirola--Kanai~\cite{MoMi-JPC-2018} and the Kostin or the Schrödinger--Langevin~\cite{MoMi-AP-2018} equations. Here, our purpose is to generalize this previous study to the von Neumann formalism of ensembles.

The last term of Equation \eqref{eq: HJ-mixed}, the quantum potential, is responsible for quantum effects. 
Following Rosen~\cite{Ro-AJP-1964}, by subtracting this term to the von Neumann equation and after again splitting the real and imaginary parts, we reach the classical Hamilton--Jacobi equation, Equation \eqref{eq: HJ-mixed}, 
without the quantum potential. Because of this, we could call this equation the {classical von Neumann equation}
(a similar classical Liouville equation could also be reached), which reads as follows
\begin{eqnarray} \label{eq: cl-vN}
i\hb \frac{\pa}{\pa t} \rho_{\cl}(x, y, t) &=& - \frac{\hb^2}{2m} 
\left( \frac{\pa^2}{ \pa x^2 } - \frac{\pa^2}{ \pa y^2 }   \right) \rho_{\cl}(x, y, t) 
+ (U(x) - U(y)) \rho_{\cl}(x, y, t)
\nonumber \\
&+& \frac{\hb^2}{2m} \left [ \frac{1}{ | \rho_{\cl}(x, y, t) |} \left( \frac{\pa^2}{ \pa x^2 } - \frac{\pa^2}{ \pa y^2 }   \right) | \rho_{\cl}(x, y, t) | \right ] \rho_{\cl}(x, y, t)  ,
\end{eqnarray}
where the sub-index ``cl'' refers to ``classical'' and $ | \rho_{\cl}(x, y, t) | $ means the modulus of 
$ \rho_{\cl}(x, y, t) $. Now, following~\cite{RiSchMaVaBa-PRA-2014}, the {transition} equation is proposed to be
\begin{eqnarray} \label{eq: tran-vN}
i\hb \frac{\pa}{\pa t} \rho_{\ep}(x, y, t) &=& - \frac{\hb^2}{2m} 
\left( \frac{\pa^2}{ \pa x^2 } - \frac{\pa^2}{ \pa y^2 }   \right) \rho_{\ep}(x, y, t) 
+ (U(x) - U(y)) \rho_{\ep}(x, y, t)
\nonumber \\
&+& (1-\ep) \frac{\hb^2}{2m} \left [ \frac{1}{ | \rho_{\ep}(x, y, t) |} \left( \frac{\pa^2}{ \pa x^2 } - \frac{\pa^2}{ \pa y^2 }   \right) | \rho_{\ep}(x, y, t) |  \right ] \rho_{\ep}(x, y, t)    .
\end{eqnarray}
%
%which contains the {transition parameter} $\ep$ going from one, for the quantum regime, to zero, for the %classical one. 
%

{From} the polar form for the density matrix 
\begin{eqnarray} \label{eq: rho-ep-polar}
\rho_{\ep}(x, y, t) &=& A_{\ep}(x, y, t) ~e^{i S_{\ep}(x, y, t) / \hb}  ,
\end{eqnarray}
{one obtains}
\begin{numcases}~
-\frac{\pa S_{\ep}}{\pa t} = \frac{ (\pa_x S_{\ep})^2 - (\pa_y S_{\ep})^2 }{ 2m } + U(x) - U(y)
- \ep \frac{\hb^2}{2m} \frac{1}{A_{\ep}} (\pa_x^2 - \pa_y^2 ) A_{\ep} 
\hspace{8pt} \label{eq: re1}
\\
\frac{\pa A_{\ep}}{\pa t} = - \frac{1}{2m} A_{\ep} (\pa_x^2 - \pa_y^2 ) S_{\ep} 
- \frac{1}{m} \left( \pa_x A_{\ep} ~ \pa_x S_{\ep} - \pa_y A_{\ep} ~ \pa_y S_{\ep} \right) 
\label{eq: im1}   .
\end{numcases}

{Now,} multiplying Equation \eqref{eq: re1} by
\begin{eqnarray} \label{eq: rhot}
\ti{\rho}(x, y, t) &=&  A_{\ep}(x, y, t) ~e^{i S_{\ep}(x, y, t) / \hbt}
\end{eqnarray}
and Equation \eqref{eq: im1} by $ i \hbt e^{i S_{\ep} / \hbt} $, and adding the resulting equations, after some straightforward algebra, one obtains
\begin{eqnarray} \label{eq: scaled-vN}
i\hbt \frac{\pa}{\pa t} \ti{\rho}(x, y, t) &=& - \frac{\hbt^2}{2m} 
\left( \frac{\pa^2}{ \pa x^2 } - \frac{\pa^2}{ \pa y^2 }   \right) \ti{\rho}(x, y, t) 
+ (U(x) - U(y)) \ti{\rho}(x, y, t)   .
\end{eqnarray}

{This} is the so-called {scaled von Neumann equation}.
The form of this scaled equation is exactly the same as that of the von Neumann equation. The only changes are that $\hb$ and $\rho$ have been replaced by the corresponding scaled quantities $ \hbt $ and $ \ti{\rho} $. 

Thus, all requirements for solutions of the conventional  Schrödinger equation (or the von Neumann 
equation in the case of mixed states) must be fulfilled here too. Based on this fact, we could
stress the following two points: (i) the solution space is formed by square-integrable functions.  
In fact, solutions must vanish at $ x \to \pm \infty $ faster than any power of $x$. This is necessary to 
have a finite value for the expectation value $ \ti{ \la x^n \ra} = \int_{-\infty}^{\infty} dx \, x^n \, \trho(x, x, t) $. 
In the example of scattering from a hard wall, which will be studied  below, at least the first two moments 
are finite; (ii) our solutions have at least two continuous derivatives, i.e, they are doubly differentiable over our domain. 
However, the first derivative of the wave function is not continuous at infinite discontinuities of the 
potential function, e.g., on hard walls where the wave function is zero.

The structure of the continuity equation is the same, and one has 
\begin{eqnarray} \label{eq: scaled-pcd}
\ti{j}(x, t) &=& \frac{\hbt}{m} \im \left\{ \frac{\pa}{\pa x} \ti{\rho}(x, y, t)\bigg|_{y=x} \right \} ,
\end{eqnarray} 
for the scaled probability density current from which the scaled velocity is derived
\begin{eqnarray} \label{eq: scaled-vel}
\ti{v}(x, t) &=& \frac{ \ti{j}(x, t) }{ \trho(x, x, t) }  .
\end{eqnarray} 

{Finally}, the scaled trajectories are determined by integrating the guidance equation
\begin{eqnarray} \label{eq: scaled-guid}
\frac{d \ti{x} }{dt} &=& \ti{v}(x, t) \bigg|_{x = \ti{x}(x^{(0)}, t)} ,
\end{eqnarray} 
with $ x^{(0)} $ being the initial position of the particle.

We now consider Ehrenfest relations. We~first write the scaled von Neumann equation~(\ref{eq: scaled-vN}) in the form
\begin{eqnarray} \label{eq: scaled-vN2}
i\hbt \frac{\pa}{\pa t} \hat{\trho} &=& [\hat{\ti{H}}, \hat{\trho}]  ,
\end{eqnarray}
where the scaled Hamiltonian operator in the position representation is 
\begin{eqnarray} \label{eq: scaled-Ham}
\ti{H} &=& -\frac{\hbt^2}{2m} \frac{\pa^2}{\pa x^2} + U(x)    .
\end{eqnarray}

{Now,} for the time derivative of the arbitrary time-independent observable $ \hat{A} $, one has~that
\begin{eqnarray} \label{eq: dAdt}
\frac{d}{dt} \wti{ \la \hat{A} \ra} &=& \tr \left ( \frac{\pa \hat{\trho} }{\pa t} \hat{A} \right )
= \frac{1}{i\hbt} \tr( [\hat{\ti{H}}, \hat{\trho}] \hat{A}  )
= \frac{1}{i\hbt} \tr( \hat{\trho} [\hat{A}, \hat{\ti{H}}] )
= \frac{ \wti{ \la [\hat{A}, \hat{\ti{H}}] \ra } }{i\hbt},
\end{eqnarray}
where we have used Equation \eqref{eq: scaled-vN2} and the cyclic property of the trace operation. Note, however, 
that one should take care of using this property when the dimension of the vector space is infinite. 
Then, from Equation \eqref{eq: dAdt}, one obtains the usual {Ehrenfest relations }
\begin{numcases}~
\frac{d}{dt} \wti{ \la \hat{x} \ra } = \wti{ \la \hat{p} \ra } \hspace{25pt}\qquad\qquad\qquad\qquad\qquad\qquad\qquad\qquad\label{eq: Eh1} \\
\frac{d}{dt} \wti{ \la \hat{p} \ra } = \wti{ \left \la - \frac{\pa U}{\pa x} \right \ra  }  . \label{eq: Eh2} 
\end{numcases}

\subsection{The Scaled Wigner--Moyal Approach}

Moyal~\cite{Mo-PCPS-1949} attempted to interpret quantum mechanics as a statistical theory. He started with 
the characteristic function, which is a standard tool of statistical theory but in the unusual way~\cite{Hi-arXiv-2014}; 
the expectation value of the so-called Heisenberg--Weyl operator~\cite{Hi-JCE-2015} was treated as the 
characteristic function. Then, the inverse Fourier transform of the characteristic function was considered to be the 
probability distribution function, and its time evolution was, thus, obtained from the standard methods of statistical 
mechanics.
{Interestingly enough, the same time evolution equation can be reached by starting from the density operator leading to
the standard quantum mechanical Liouville--von Neumann equation. }
In spite of seemingly different starting points, Hiley~\cite{Hi-arXiv-2014} has shown they are, in fact, 
the same starting point. We~are going to follow the same procedure but in the scaled theory context.
%Here, we use this second approach to present our scaled theory within the context of Wigner-Moyal for pure states.

Using the Fourier transform of the scaled wavefunction, the corresponding pure scaled density matrix can be written as
\begin{eqnarray}
 \trho(x, y, t) &=& \tpsi(x,t) \tpsi^*(y,t)  = \frac{1}{2 \pi \hbt} \int dp dq ~ \tphi(p,t) ~ \tphi^*(q,t)  ~ e^{i (p x - q y ) / \hbt} ,
\end{eqnarray}
and, again using the $R$ and $r$ coordinates for space and, similarly, $ u = (p+q)/2 $ and $ v = p-q $ 
for momentum, the density matrix can be transformed into
\begin{eqnarray}
\ti{\rho}(R, r, t) &=& \frac{1}{2 \pi \hbt} \int du dv~ \ti \phi(u+v/2,t) \ti \phi^* (u-v/2,t) e^{i (R v + r u) / \hbt } \nonumber \\
& \equiv & \int du ~ \tiW(R,u,t) e^{i u r / \hbt}   \label{eq: rho-F}.
\end{eqnarray}
%
%where we have defined 
%
%\begin{eqnarray} \label{F}
%\tiF(R,u,t) &=& \frac{1}{2 \pi \hbt} \int dv~ \ti \phi(u+v/2,t) \ti \phi^* (u-v/2,t) e^{i R v / \hbt }.
%\end{eqnarray}
%

{This} equation shows that the function $ \tiW(R,u,t) $ is the partial Fourier transform of the scaled density matrix $ \trho(R,r,t) $ with respect to the relative coordinate $ r $. Thus, one has that
\begin{eqnarray}
\tiW(R,u,t) &=& \frac{1}{2 \pi \hbt} \int dr ~ e^{-i u r / \hbt} \trho(R,r,t)  \label{eq: F-rho} \\
&=& \frac{1}{2 \pi \hbt} \int dr ~ e^{-i u r / \hbt} \tpsi(R+r/2, t) \tpsi^*(R-r/2, t)  , \nonumber 
%\\&=& \frac{1}{\pi \hbt} \int dr' ~ e^{i u r' / \hbt} \tpsi(R-r', t) \tpsi^*(R+r', t) \nonumber
\end{eqnarray}
which is just the {scaled} Wigner distribution function (see Ref.~\cite{Wigner} for comparison). 
This can be explicitly seen by changing the relative variable $ r \to - r/2 $. 
The time evolution of the scaled Wigner distribution function $ \tiW(R,u,t) $ can be found from 
Equations (\ref{eq: F-rho}) and (\ref{eq: scaled-vN}), written in the coordinates $r$ and $R$, according to
\begin{eqnarray} \label{eq: dFdt}
\frac{\pa}{\pa t}\tiW(R, u, t) &=& \frac{1}{2 \pi \hbt} \int dr ~ e^{-i u r / \hbt} \frac{\pa}{\pa t} \trho(R,r,t)  \nonumber \\  
&=& \frac{1}{2 \pi \hbt} \int dr ~ e^{-i u r / \hbt} \left( \frac{i\hbt}{m} \frac{\pa^2}{\pa r \pa R}
+ \frac{ U(R/2+r) - U(R/2-r) }{i\hbt} \right) \trho(R, r, t) \nonumber \\
&=& - \frac{u}{m}  \frac{\pa}{\pa R} \tiW(R, u, t) 
+ \int du' \ti{K}(R, u'-u, t) \tiW(R, u',t)    ,
\end{eqnarray}
where we have defined the kernel 
\begin{eqnarray} \label{eq: J-kernel}
\ti{K}(R, q,t) &=& \frac{1}{2 \pi \hbt} \int dr ~ e^{i q r / \hbt} \frac{ U(R/2+r) - U(R/2-r) }{i\hbt} .
\end{eqnarray}

{Thus}, one finally has that 
\begin{equation}
\frac{\partial \ti W(x, p, t)}{\partial t} + \frac{p}{m} \frac{\partial \ti W(x,p,t)}{\partial x} = \int \ti{K}(x,p-q) \ti W(x,q,t) dq	.
\end{equation}	

In the so-called Wigner--Moyal approach to quantum mechanics and, as shown before, Moyal's starting point 
\cite{Mo-PCPS-1949} is the Heisenberg--Weyl operator, which is defined as 
\begin{equation} \label{eq: M-op}
\hat{M}(\theta, \uptau) = e^{i (\theta \hat x+\uptau \hat p)}  = e^{i\uptau \hat p / 2 } e^{i \theta \hat x } e^{ i \uptau \hat p / 2 }  ,
\end{equation}
and its expectation value, considered as the characteristic function, is
\begin{equation}
M(\theta, \uptau) = \int dx~ \psi^*(x) e^{i\uptau \hat p / 2 } e^{i \theta \hat x } e^{ i \uptau \hat p / 2 } \psi(x) .
\end{equation}

Now, the same procedure could be followed in this context and written as
\begin{eqnarray} \label{eq: tiM}
\ti M(\theta, \uptau) &=& \int dx ~ \ti \psi^*(x) e^{i\uptau \hat p / 2 } e^{i \theta \hat x } e^{ i \uptau \hat p / 2 } \tpsi(x)  \nonumber \\
&=& \int dx ~ \tpsi^*(x-\hbt \uptau/2) ~ e^{i \theta x} ~ \tpsi(x+\hbt \uptau/2)  ,
\end{eqnarray}
wherein the phase space probability distribution function is the Fourier transform of the characteristic function
\begin{eqnarray}
\ti W(x,p) &=& \frac{1}{(2\pi)^2} \int \int d\uptau ~ d\theta ~\ti M(\theta, \uptau) e^{ - i (\theta x+\uptau p)}  \nonumber \\
&=& \frac{1}{2\pi} \int d\uptau ~\tpsi^*(x-\hbt \uptau/2) ~ e^{-i \uptau p} ~ \tpsi(x+\hbt \uptau/2) .
\end{eqnarray}

{In the} second line of Equation \eqref{eq: tiM}, we have used the fact that the momentum operator is the generator 
of translations. As Moyal proposed, one could also consider $ \ti W(x,p) $ as a distribution function and apply 
the corresponding standard methods of mechanical statistics. 
%In particular, the time derivative of the corresponding characteristic function assumed to satisfy a quantum Liouville equation 
%
%\begin{equation}
%i \ti \hbar \frac{\partial \hat M}{\partial t} = [\ti H, \hat M]	,
%\end{equation}
% 
%with $\ti H = - \frac{- \ti \hbar^2}{2 m} \frac{\partial^2}{\partial x^2} + U(x)$. 
%
Starting from the Heisenberg equation of motion
\begin{eqnarray} \label{eq: Heis}
\frac{d}{dt} \thM &=& \frac{ [\thM, \thH]  }{ i \hbt }, ~~~~~~~ \ti{H} = - \frac{ \ti \hbar^2}{2 m} \frac{\partial^2}{\partial x^2} + U(x)
\end{eqnarray}
for the scaled operator $ \thM $, and following Moyal's original work~\cite{Mo-PCPS-1949}, one anticipates
\begin{eqnarray} \label{eq: F-evol}
\frac{\pa}{\pa t} \tiW(x, p, t) &=& \frac{2}{\hbt} 
\sin\left[ \frac{\hbt}{2} \left( \frac{\pa}{\pa p_{\tiW}} \frac{\pa}{\pa x_H} 
- \frac{\pa}{\pa p_H } \frac{\pa}{\pa x_{\tiW}}
\right) \right] H(x, p) \tiW(x, p, t)   ,
\end{eqnarray}
where $ H(x, p) $ is the classical Hamiltonian, and $ \pa/ \pa x_{\tiW} $ and $ \pa/ \pa p_{\tiW} $ operate only on $  \tiW $ and so forth.
In the classical limit $ \ep \to 0 $, this equation reduces to the well-known Liouville equation for the phase 
space distribution function,
\begin{eqnarray} \label{eq: Lio}
\frac{\pa}{\pa t} W_{\cl}(p, q, t) &=& \{ W_{\cl}, H \}_{\text{PB}}   ,
\end{eqnarray}
where $ \{ \cdot, \cdot \}_{\text{PB}} $ stands for the Poisson bracket.

Thus, we have built a scaled nonequilibrium statistical mechanics equation %Please check meaning retained
 from its fundamentals, 
which takes into account, in a continuous and smooth way, all the dynamical regimes in-between the 
two extreme case, as well as the quantum and classical ones.

\section{Results and Discussion}

As a simple application of our theoretical formalism, let us consider scattering from a hard wall at the origin
\begin{eqnarray} \label{eq: hardwall}
V(x) &=& 
\begin{cases}
0 & x <0 \\
\infty & 0 \leq x   ,
\end{cases}
\end{eqnarray}
and two scaled Gaussian wave packets $ \tpsi_a $ and $ \tpsi_b $ with the same width $ \si_0 $ but different centers 
$ x_{0a} $ and $ x_{0b} $ (initially localized in the left side of the wall) and  kick momenta $ p_{0a} $ and $ p_{0b} $, i.e.,
\begin{numcases}~
	\tpsi_a(x, 0) = \frac{1}{ (2\pi \si_0^2)^{1/4} } \exp \left[ -\frac{ (x - x_{0a})^2 }{ 4\si_0^2 } + i \frac{p_{0a}}{\hbt}x \right] 
	\hspace{18pt}\qquad\qquad\qquad\\
	\tpsi_b(x, 0) = \frac{1}{ (2\pi \si_0^2)^{1/4} } \exp \left[ -\frac{ (x - x_{0b})^2 }{ 4\si_0^2 } + i \frac{p_{0b}}{\hbt}x \right] .
\end{numcases}
We build the superposition state at any time as
\begin{eqnarray} \label{eq: purestate}
\tpsi(x, t) &=& \ti{ \mathcal{N} } \frac{ \tpsi_a(x, t) + \tpsi_b(x, t) }{\sqrt{2}} \theta(-x)   
\end{eqnarray}
and the corresponding mixture as
\begin{eqnarray} \label{eq: mixedstate}
\trho(x, y, t) = \frac{\tpsi_a(x, t) \tpsi_a^*(y, t) + \tpsi_b(x,t) \tpsi_b^*(y, t)}{2}  \theta(-x)   ,
\end{eqnarray}
where $\theta(x)$ is the step function, and $ \ti{ \mathcal{N} } $ is the normalization constant. 
Note that the unitary evolution of the scaled wave functions under the corresponding  von Neumann 
equation keeps the purity of states, which is quantified from $ \tr( \ti{\hat{\rho}}^2 ) $. By using now the propagator for the hard wall potential~\cite{GrSt-book-1998} 
\begin{eqnarray} \label{eq: hw_prop}
\ti{G}(x, t; x', 0) &=& \ti{G}_{\free}(x, t; x', 0) - \ti{G}_{\free}(-x, t; x', 0)  ,
\end{eqnarray}
one obtains  that
\begin{eqnarray} \label{eq: hw_wf}
\tpsi(x, t) &=& ( \tpsi_{\free}(x, t) - \tpsi_{\free}(-x, t) )  \theta(-x)  ,
\end{eqnarray}
where the sub-index ``f'' stands for ``free``, and the corresponding propagator is written as
\begin{eqnarray} \label{eq: free-prop}
\ti{G}_{\free}(x, t; x', 0) &=& \sqrt{ \frac{m}{ 2\pi i \hbt t} } \exp \left[ \frac{i m}{2 \hbt t} (x-x')^2 \right]   .
\end{eqnarray}

{Now}, from Equation \eqref{eq: hw_wf}, one reaches
\begin{eqnarray}
\tpsi_a(x, t) &=& \bigg \{ \frac{1}{ (2\pi \tst^2)^{1/4} } \exp \left[ -\frac{ (x - x_{ta})^2 }{ 4\si_0 \tst } + i \frac{p_{0a}}{\hbt}(x - x_{ta}) + i \frac{p_{0a}^2 t}{2m\hbt} + i \frac{p_{0a} x_{0a}}{\hbt} \right] 
\nonumber \\
&+& 
\frac{1}{ (2\pi \tst^2)^{1/4} } \exp \left[ -\frac{ (x + x_{ta})^2 }{ 4\si_0 \tst } - i \frac{p_{0a}}{\hbt}(x + x_{ta}) + i \frac{p_{0a}^2 t}{2m\hbt} + i \frac{p_{0a} x_{0a}}{\hbt} \right] 
 \bigg \} \theta(-x)  , \nonumber \\
\end{eqnarray}
{where}%MDPI:subequations are not allowed, pelase check if content below coul dbe marked as Equation (78).
\begin{numcases}~
\tst = \si_0 \left( 1 + i \frac{\hbt t}{2 m \si_0^2} \right) \label{eq: tst} \\
x_{ta} = x_{0a} + \frac{ p_{0a} }{m} t   , \label{eq: xta}
\end{numcases}
being the complex width and the center of the freely propagating Gaussian wavepacket, respectively. 
The same holds for the $b$ component of the wave function when replacing $a$~by~``$b$''.

%=========================================

\begin{figure}
	\includegraphics[width=12cm,angle=-0]{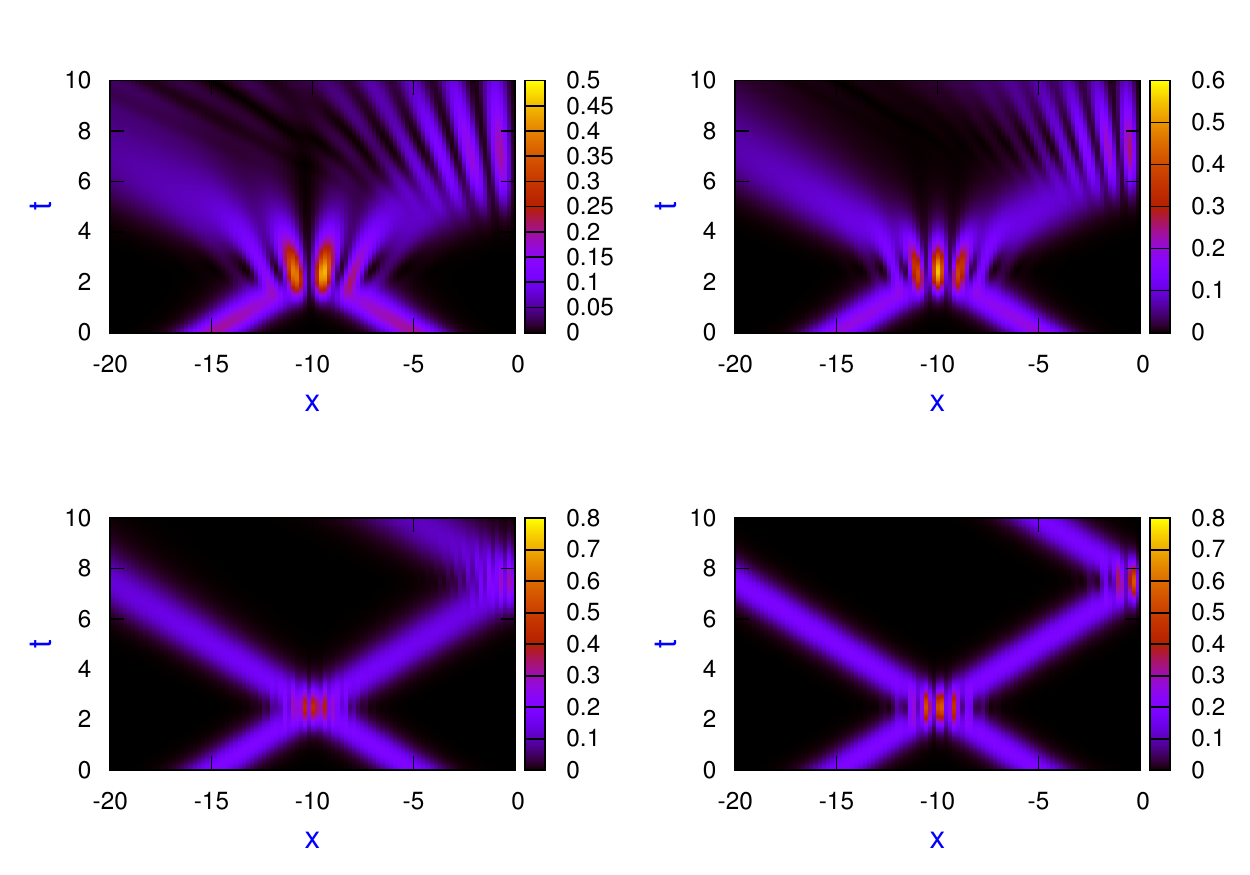}
	\caption{
		{Scaled probability} density plots for the superposition of two Gaussian wave packets, Equation \eqref{eq: purestate}, for different regimes: 
		$ \ep = 1 $ (left top panel), $ \ep = 0.5 $ (\textbf{right top panel}), $ \ep = 0.1 $ (\textbf{left bottom panel}), and $ \ep = 0.01 $ (\textbf{right bottom panel}).
		We used as initial parameters, $ m = 1 $, $ \hb = 1 $, $ p_{0b} = - p_{0a} = 2 $, 
		$ \si_{0b} = \si_{0a} = 1 $, $ x_{0a} = -5 $ and $ x_{0b} = -15 $.
	} 
	\label{fig: denprob_pure} 
\end{figure}

\begin{figure}
	\includegraphics[width=12cm,angle=-0]{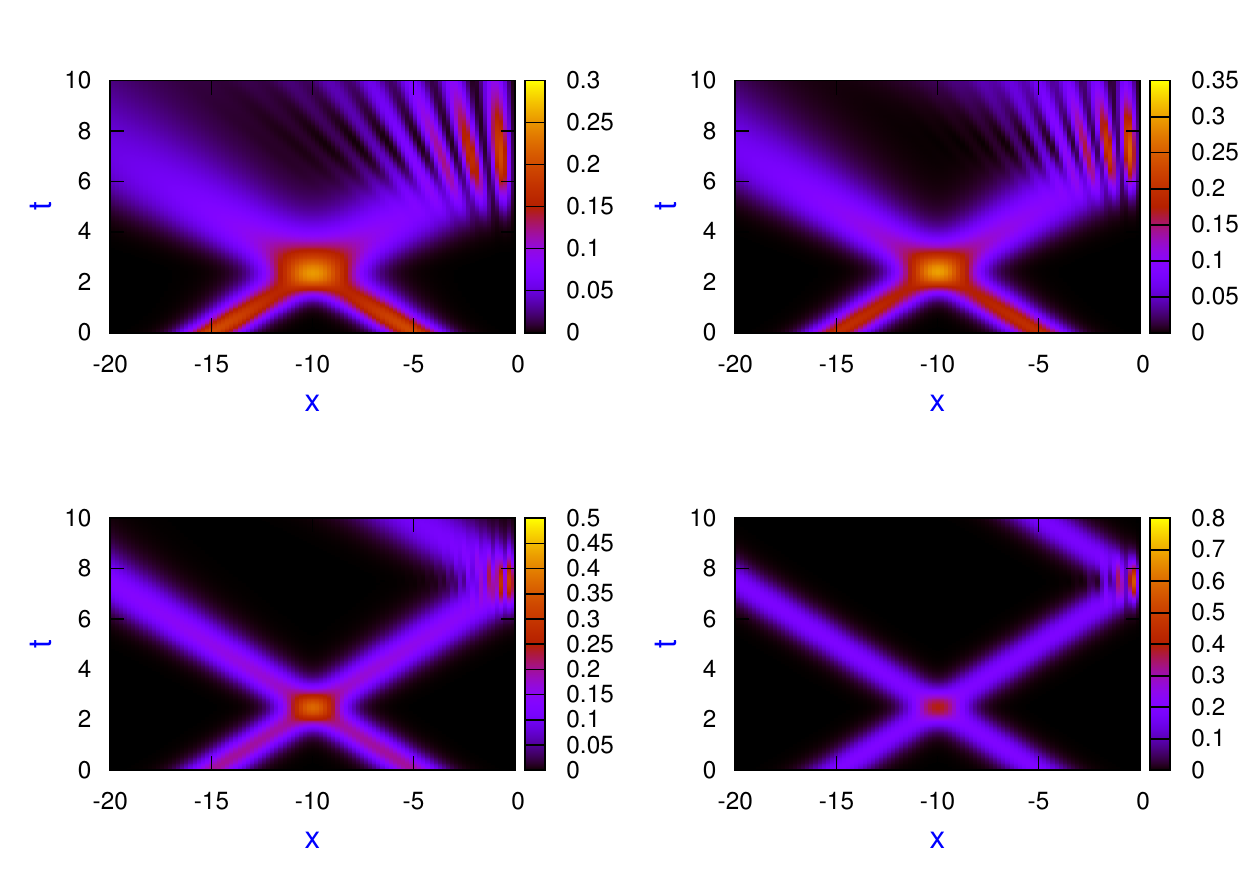}
	\caption{
		{Scaled probability} density plots for the mixed state, Equation \eqref{eq: mixedstate}, for different regimes: 
		$ \ep = 1 $ (left top panel), $ \ep = 0.5 $ (\textbf{right top panel}), $ \ep = 0.1 $ (\textbf{left bottom panel}), and $ \ep = 0.01 $ (\textbf{right bottom panel}).
		%
		%We have again used $ m = 1 $, $ \hb = 1 $, $ p_{0b} = - p_{0a} = 2 $, $ \si_{0b} = \si_{0a} = 1 $, $ x_{0a} = -5 $ and $ x_{0b} = -15 $.
		The same initial parameters as Figure~\ref{fig: denprob_pure} were used. 
	} 
	\label{fig: denprob_mixed} 
\end{figure}

In Figures~\ref{fig: denprob_pure} and~\ref{fig: denprob_mixed}, scaled {probability density plots for the superposition 
of two Gaussian wave packets}, Equation \eqref{eq: purestate}, and for the mixed state, Equation \eqref{eq: mixedstate}, for 
different dynamical regimes are shown: $ \ep = 1 $ (left top panel), $ \ep = 0.5 $ (right top panel), $ \ep = 0.1 $ 
(left bottom panel), and $ \ep = 0.01 $ (right bottom panel). In ~both figures, the following initial parameters were used for the calculations: $ m = 1 $, $ \hb = 1 $, $ p_{0b} = - p_{0a} = 2 $, $ \si_{0b} = \si_{0a} = 1 $, $ x_{0a} = -5 $, and $ x_{0b} = -15 $. As can clearly be seen in both cases, when the transition parameter was approaching zero (the classical dynamics
regime), the interference pattern in collision between both states ,as well as in the scattering from the hard wall, tended to be 
washed up in a continuous way.   As one expects, when approaching  the classical regime, results for the pure and the mixed states also became closer. 

\begin{figure}
	\includegraphics[width=12cm,angle=-0]{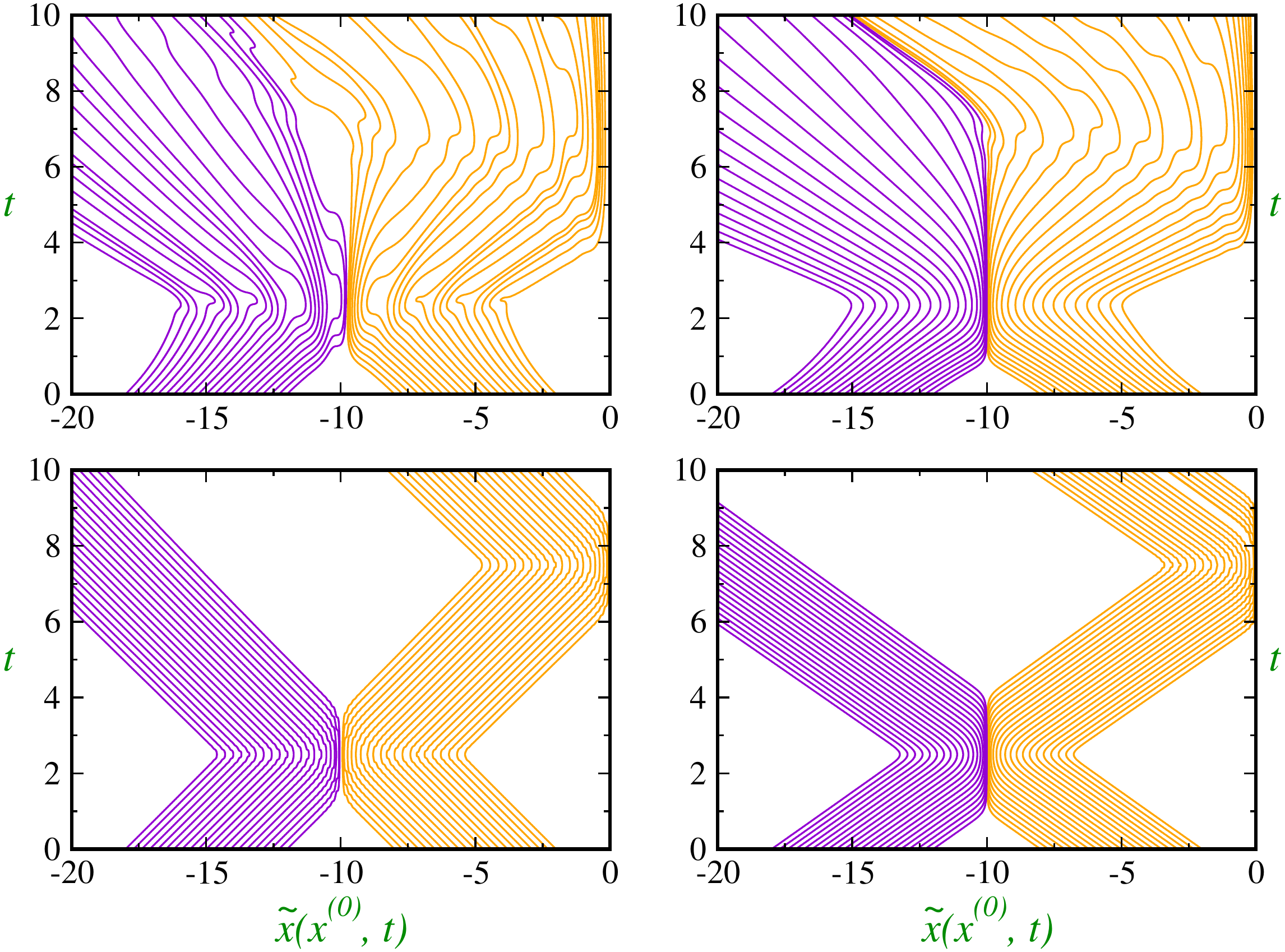}
	\caption{
		{A selection }of scaled trajectories for the scaled pure $ \ti{\psi}(x, t)$ (\textbf{left column}) and the scaled mixed state 
		$ \ti{\rho}(x, y, t) $ (\textbf{right column}) for the quantum regime $\ep = 1$ \textbf{(top panels}) and the  nearly classical 
		regime $ \ep = 0.01 $ (\textbf{bottom panels}). The same  initial parameters were used as in previous figures.
	} 
	\label{fig: trajs} 
\end{figure}

Let us discuss now how the {scaled} trajectories behaved in the different dynamical regimes, going from 
Bohmian trajectories ($\ep = 1$) to pure classical ones ($\ep = 0$).
In Figure~\ref{fig: trajs},  a selection of scaled trajectories is plotted for the scaled wave function $ \ti{\psi}(x, t) $ (left column) and the scaled density matrix $ \ti{\rho}(x, y, t) $ (right column) for the quantum regime (top panels) and the  nearly classical dynamical regime $ \ep = 0.01 $ (bottom panels). The same units and initial parameters were used as previously. 
Comparison with the Figures~\ref{fig: denprob_pure} and~\ref{fig: denprob_mixed} reveals that trajectories followed the wave packets. In ~addition, although it is not apparent from our figure, if one had selected the distribution of the initial positions according to the Born rule then he/she would have seen compact trajectories in regions with higher values of probability distribution. In ~other words, if trajectories obey the Born rule initially, they will do so forever.
The non-crossing rule of trajectories was still observed at the nearly classical regime $ \ep = 0.01 $ and even in the classical regime, which is a consequence of the {first order classical theory} in contrast to the {true second order theory}.  
As the classical regime was approaching, the corresponding trajectories became more localized by simulating two
classical collisions---with the first one coming from the scattering between the two wave packets and the second from 
the wall. However, only the wave packet starting closer to the hard wall was reflected by the wall due to the 
second collision. 
As has also been discussed elsewhere~\cite{Salva2008}, wave packet interference can also be 
understood within the context of scattering off effective potential barriers. In~classical mechanics, one can 
always substitute a particle--particle collision by that of an effective particle interacting with a potential. This fact is clearly observed in this context both for the superposition wave packet as well as for the density matrix. 

From the non-crossing property of  Bohmian trajectories, Leavens~\cite{Le-book-2008}  proved that the 
arrival time distribution is given by the modulus of the probability current density. 

{Following t}he same procedure for  
the  scaled trajectories, one has that the {scaled arrival time distribution} at the detector place $X$ can be 
expressed as 
\begin{eqnarray} \label{eq: ardis}
 \ti{\Pi}(X, t) = \frac{| \ti{j}(X, t) |}{ \int_0^{\infty} dt' | \ti{j}(X, t') | }   .
\end{eqnarray}

{Moreover}, the mean arrival time at the detector location and the variance in the measurement of the arrival time, which 
is also a measure of the width of the distribution, are respectively given by  
\begin{eqnarray} 
\wti{ \la t \ra } &=& \int_0^{\infty} dt'~t'~ \ti{\Pi}(X, t') \label{eq: meanar} \\
\wti{ \Delta t } &=& \sqrt{ \wti{ \la t^2 \ra } - \wti{ \la t \ra }^2 }    .  \label{eq: varar} 
\end{eqnarray}

{As a result }of the previous analysis in terms of scaled trajectories, these quantities are easily calculated.

\begin{figure}
	\includegraphics[width=12cm,angle=-0]{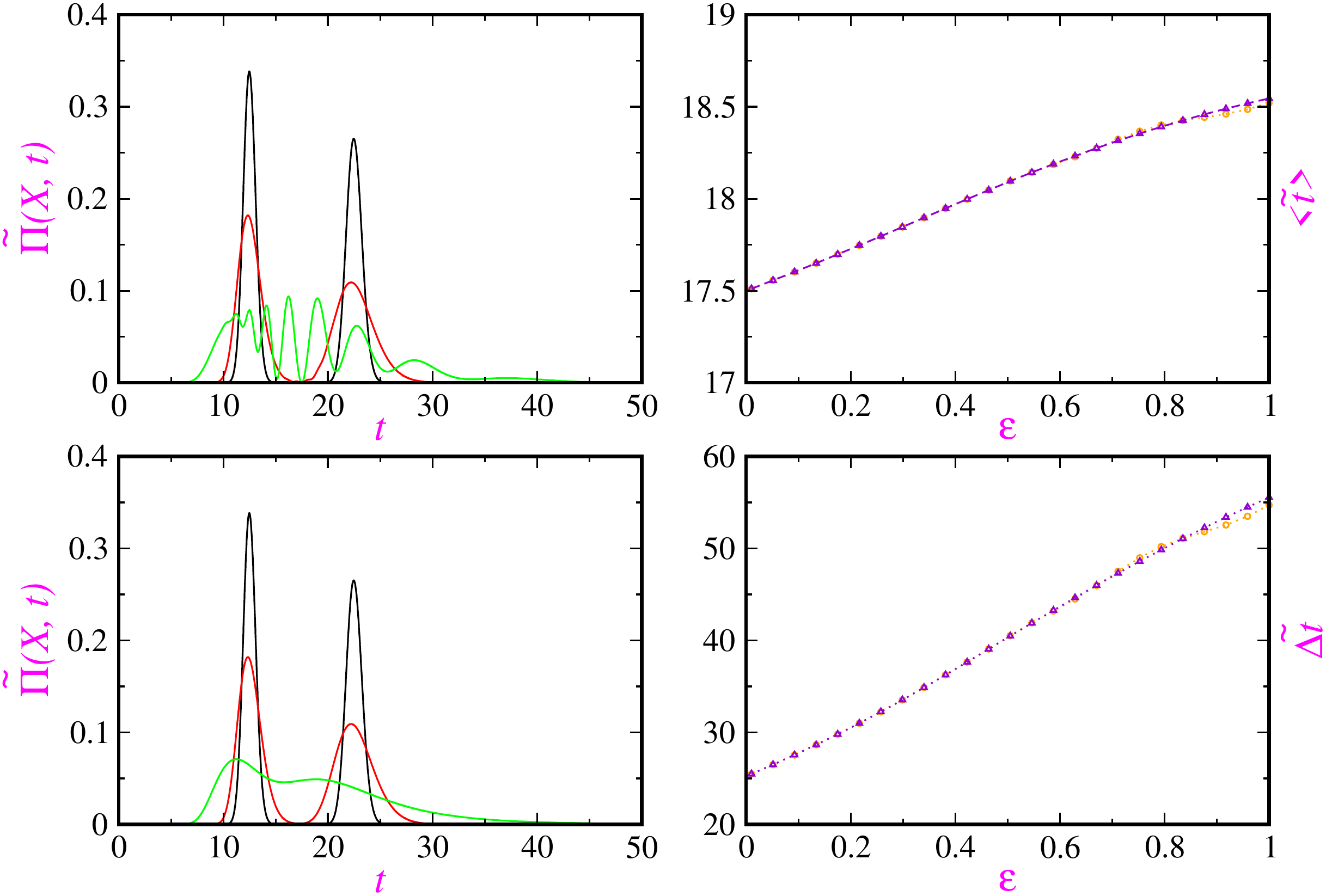}
	\caption{
		Scaled arrival time distribution (\ref{eq: ardis}) at the detector location $ X = -30 $ for the pure state (\ref{eq: purestate}) (\textbf{left top panel}) and the mixed state (\ref{eq: mixedstate}) (\textbf{left bottom panel}) for different regimes: 
		$ \ep = 1 $ (green curve), $ \ep = 0.1 $ (red curve), and $ \ep = 0.01 $ (black curve).
		{Right top} (\textbf{bottom}) panel depicts the scaled mean (uncertainty in) arrival time versus the transition parameter for the pure state (orange circled) and the mixed state (violet triangle up). The same initial parameters were used as in previous figures.
} 
	\label{fig: ar_p0b=-p0a=2} 
\end{figure}

In Figure~\ref{fig: ar_p0b=-p0a=2}, scaled arrival time distributions  have been plotted at the detector location $ X = -30 $ for the pure state (\ref{eq: purestate}) (left top panel) and the mixed state (\ref{eq: mixedstate}) (left bottom panel) for three different dynamical regimes: 
$ \ep = 1 $ (green curve), $ \ep = 0.1 $ (red curve), and $ \ep = 0.01 $ (black curve).
On the right top and bottom panels,  the scaled mean arrival time and variance versus the transition parameter for the pure state (orange circled) and the mixed state (violet triangle up) have been also displayed.
As clearly seen, the mean arrival time diminished when going from the quantum to classical regime. This is related to the width of the probability distribution, which was wider for the quantum regime than for the classical one.
%In addition, the width of the arrival time decreases in transition from quantum to classical regime. 
Furthermore, some differences between results coming from for the pure and mixed states seemed to appear only around the quantum regime.

\begin{figure}
	\includegraphics[width=12cm,angle=-0]{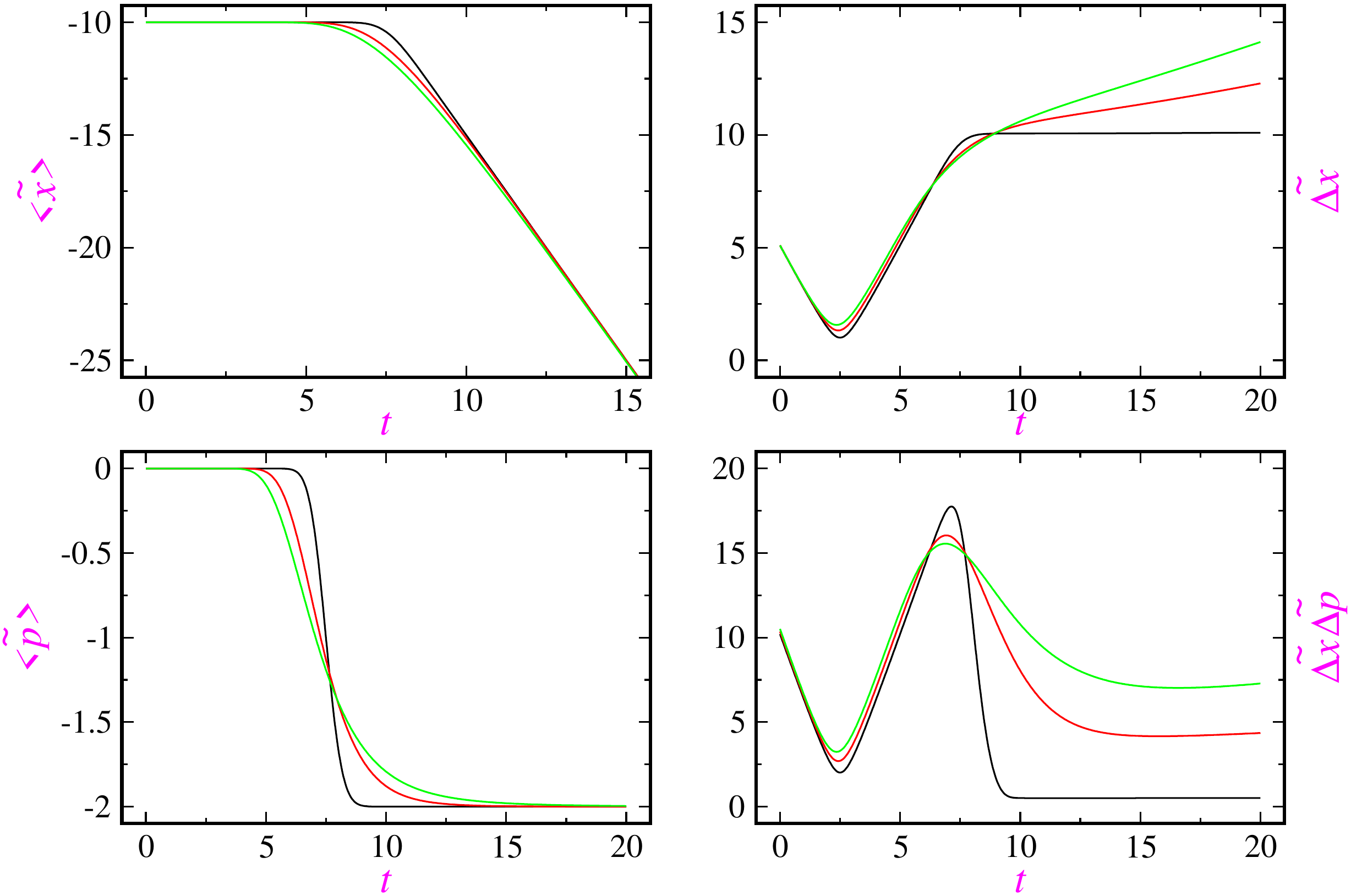}
	\caption{
	{	Expectation }value of position operator (\textbf{left top panel}), uncertainty in position (\textbf{right top panel}), expectation 
		value of momentum operator (\textbf{left bottom panel}), and the product of uncertainties (\textbf{right bottom panel}) for the scaled mixed state 
		$ \ti{\rho}(x, y, t) $ for three different dynamical regimes: $ \ep = 1 $ (green curves), $ \ep = 0.5 $ (red curves), 
		and $ \ep = 0.01 $ (black curves). The same initial parameters were used as in previous figures.
	} 
	\label{fig: expectations_mixed_p0b=-p0a=2} 
\end{figure}

In Figure~\ref{fig: expectations_mixed_p0b=-p0a=2}, the expectation value of position operator (left top panel),
the uncertainty in position (right top panel), the expectation value of the momentum operator (left bottom panel), and the product of uncertainties for the scaled mixed state $ \ti{\rho}(x, y, t) $ for three different dynamical regimes are shown: 
$ \ep = 1 $ (green curves), $ \ep = 0.5 $ (red curves), and $ \ep = 0.01 $ (black curves) are plotted. The same  
initial parameters were used as in previous figures.
This figure shows that  the continuous transition from the quantum to classical dynamical regime presented 
several global and important features: 
(i) reflection from the wall  was delayed on average; (ii) the average velocity in reflection decreased; 
(iii) the uncertainty in position, which is also a measure of width of the state, diminished; and 
(iv) the product of uncertainties also decreased at long times.
Furthermore, the  scaled Heisenberg uncertainty relation $ \wti{\Delta x} \wti{\Delta p} \geq \hbt/2  = \sqrt{\ep} \hb / 2 $ holding in any dynamical regime can be proved straightforwardly starting from the definition of uncertainties and the scaled von Neumann equation; this equation has the same mathematical form as the usual one and, thus, everything goes fine. 
This relation, holding separately in all dynamical regimes, does not compare different regimes with each other. 
The bottom right panel of Figure~\ref{fig: expectations_mixed_p0b=-p0a=2} confirms the fulfilment of this relation for all regimes. 
There was a time interval around $ t \approx 7 $ for which the product of uncertainties increased 
when approaching to the classical regime. According to Figures~\ref{fig: denprob_mixed}, this time interval corresponds to the reflection time from the wall.

Numerical calculations showed that, during reflection from the wall, both uncertainties became higher when approaching the classical regime.  
A rough explanation for this seemingly unexpected result is the following.
As top panels of Figure~\ref{fig: denprob_mixed} show, the interference pattern structure (light and 
dark regions) was seen during the reflection, around $ t \approx 7 $, from the wall for the quantum and 
nearby regimes, while the pattern was more or less smooth in the classical and nearby regimes. In ~dark 
regions, $ \trho \approx 0 $, and, thus, these regions did not have contributions to uncertainty, which itself 
is a measure of width of the distribution. Therefore, this width having some contribution only from light 
regions increased in the reflection time for the classical regime in comparison with the quantum~one.

An interesting quantity is the {non-classical effective force}, which is defined via
\begin{eqnarray}
\wti{ f_{\nc} } &=& \frac{d}{dt} \wti{ \la \hat{p} \ra }   .
\end{eqnarray}

{From} the mixture (\ref{eq: mixedstate}), one has that
\begin{eqnarray} \label{eq: mom-exp-mix}
\wti{ \la \hat{p} \ra } &=& \frac{1}{2} ( \wti{ \la \hat{p} \ra_a } + \wti{ \la \hat{p} \ra_b } )  ,
\end{eqnarray}
where $ \wti{ \la \hat{p} \ra_i } $ is the expectation value of the momentum operator with respect to the component wavefunction $ \tpsi_i(x ,t) $. From~the scaled  Schrödinger equation (\ref{eq: Scaled Sch}) and boundary conditions on the wavefunction and its space derivative, one obtains
\begin{eqnarray} \label{eq: dp-dt-pure}
\frac{d}{dt} \wti{ \la \hat{p} \ra_i } &=& - \frac{\hbt^2}{2m} \left| \frac{\pa \tpsi_i}{\pa x} \right|^2 \bigg|_{x=0}.
\end{eqnarray}

{Finally}, from Equations (\ref{eq: mom-exp-mix}) and (\ref{eq: dp-dt-pure}), one has that
\begin{eqnarray}
\wti{ f_{\nc} } &=& - \frac{1}{2} \frac{\hbt^2}{2m} \left( \left| \frac{\pa \tpsi_a}{\pa x} \right|^2 + \left| \frac{\pa \tpsi_b}{\pa x} \right|^2 \right) \bigg|_{x=0}.
\end{eqnarray}

{{Classically}}, there is no force in the region $ x < 0 $. In ~this regime, particles' momentums reverse suddenly 
at the collision time with the hard wall; however, this is not the non-classical case, as the left-bottom panel of 
Figure~\ref{fig: expectations_mixed_p0b=-p0a=2} shows. Only classical particles with initial positive momentums, 
(in our case, particles described by $\tpsi_b$) collide with the wall which, for our initial parameters, the collision time was 
$ m x_{0b} / p_{0b} = 7.5 $.

\section{Discussions and Conclusions}

Along the last years, we have shown that scaled trajectories provide an alternative and complementary view of the 
so-called quantum-to-classical (smooth) transition within a theoretical scheme that is similar to the well-known WKB approximation. The (internal) decoherence process is also well and continuously established when
approaching the classical limit. 
Note that we used ``classical limit'' in the sense of ``classical-like behavior``, and this does not necessarily 
imply that the system of interest is macroscopic. 
The tunnelling effect, as well as the diffusion problem within the 
Langevin framework, were successfully applied.
In~this work, we extended this theoretical formalism  in the same direction to propose a scaled Liouville--von Neumann equation and
its Wigner representation, which is precisely the first step to build 
a scaled nonequilibrium statistical mechanics function. %Please check meaning retained 
%, starting from a scaled Schrödinger equation,
This was carried out following the same procedure proposed 
by Moyal a long time ago in the context of standard quantum mechanics. This approach opens 
up new avenues to develop that consist of, for example, a scaled Fokker--Planck equation, 
the analysis of phase transitions, space-time correlation functions, master equations, reaction rates, and 
the so-called Kramers' problem, including kinetic models, linear response theory, projection 
operators, mode-coupling theory,  non-linear transport equations,  and much more. For example, 
the book by Zwanzig~\cite{Zwanzig-2001} could be a good guide to follow in the near future. 
%Work in this direction is now in progress.
Note also that  our scaled approach, in this context, was different from the environmental decoherence 
or collapse models such as the well-known Ghirardi--Rimini--Weber model. Here, the goal was not to find 
an explanation for the reduction postulatem but the proposal for a scaled approach, which looks for 
describing a smooth transition from quantum-to-classical mechanics in terms of a wave function or
density matrix. 
For a discussion on the measurement within the classical  Schrödinger equation see, for example,
Ref.~\cite{Ni-FPL-2006}, where, by formulating the measurement theory in this non-linear context, it was 
argued that all {measurable} properties of classical mechanics can be predicted by the quantum theory 
based on the classical Schrödinger equation, without assuming the existence of particle trajectories.


\begin{thebibliography}{999}
%
\bibitem{AnAh-FPL-1999}
Anandan, J.; Aharonov, Y. Meaning of the density matrix. \emph{Found. Phys. Lett.}  \textbf{1999}, \emph{12}, 571.
%
\bibitem{Ma-FP-2005}
Maroney, O.J.E. The Density Matrix in the de Broglie-Bohm Approach. \emph{Found. Phys.} \textbf{2005}, \emph{35}, 493.
%
\bibitem{DuGoTuZa-FP-2005}
D\`urr, D.;Goldstein, S.;Tumulka, R.; Zangh, N.I. On the role of density matrices in Bohmian mechanics. \emph{Found. Phys.} \textbf{2005}, \emph{35}, 449.
%
%
\bibitem{JoZeKiGiKuSt-book-2002} 
Joos, E.; Zeh, H.D.; Kiefer, C.; Giulini, D.; Kupsch, J.;  Stamatescu, I.-O. {\it Decoherence and the Appearance of a Classical World in Quantum Theory}; Springer: Berlin/Heidelberg, Germany, 2002.
%
\bibitem{Sch-book-2007} 
Schlosshauer, M. {\it Decoherence and the Quantum-to-Classical Transition}; Springer: Berlin/Heidelberg, Germany, 2007.
%
\bibitem{Zu-PT-1991} 
Zurek, W.H.  Decoherence and the transition from quantum to classical. \emph{Phys. Today} \textbf{1991}, \emph{44}, 36.
%
\bibitem{Mi-PRA-1991}
Milburn, G.J. Intrinsic decoherence in quantum mechanics. \emph{Phys. Rev. A} \textbf{1991},  \emph{44}, 5401. 
%
\bibitem{GRW-PRD-1986}
Ghirardi, G.C.; Rimini, A.; Weber, T. Unified dynamics for microscopic and macroscopic systems. \emph{Phys. Rev. D} \textbf{1986}, \emph{34}, 470. 
%
%
\bibitem{Holland-book-1993} 
Holland, P.R. {\it The Quantum Theory of Motion}; Cambridge University Press: Cambridge, MA, USA, 1993. 
%
\bibitem{salva1} 
Sanz, A.S.; Miret-Art\'es, S. {\it A Trajectory Description of Quantum Processes. Part I. Fundamentals}; Lecture Notes in Physics; Springer: Berlin/Heidelberg, Germany,  2012; Volume 850.
%
\bibitem{salva2} 
Sanz, A.S.; Miret-Art\'es, S.  {\it A Trajectory Description of Quantum Processes. Part II. Applications}; Lecture Notes in Physics; Springer: Berlin/Heidelberg, Germany, 2014; Volume 831. 
%
%
\bibitem{NaMi-book-2017}
Nassar, A.B.; Miret-Art\'es, S. {\it Bohmian Mechanics, Open Quantum Systems and Continuous
Measurements}; Springer: Berlin/Heidelberg, Germany, 2017.
%
\bibitem{RiSchMaVaBa-PRA-2014}
Richardson, C.D.; Schlagheck, P.; Martin, J.; Vandewalle, N.;  Bastin, T. Nonlinear Schr\"odinger wave equation with linear quantum behavior. \emph{Phys. Rev. A} \textbf{2014}, \emph{89}, 032118.
%
\bibitem{MoMi-AP-2018} 
Mousavi, S.V.; Miret-Art\'es, S. Dissipative tunnelling by means of scaled trajectories. \emph{Ann. Phys.} \textbf{2018}, \emph{393}, 76.
%
\bibitem{MoMi-JPC-2018}
Mousavi, S.V.; Miret-Art\'es, S. Quantum-classical transition in dissipative systems through scaled trajectories. \emph{J. Phys. Commun.} \textbf{2018}, \emph{2}, 35029.
%
\bibitem{Chou1-2016}
Chou, C.-C. Trajectory description of the quantum-classical transition for wave packet interference. \emph{Ann. Phys.} \textbf{2016}, \emph{371}, 437.
%
\bibitem{Chou2-2016}
Chou, C.-C. Quantum‐classical transition equation with complex trajectories. \emph{Int. J. Quan. Chem.} \textbf{2016}, \emph{116}, 1752.
%
\bibitem{MoMi-FOP-2022}
Mousavi, S.V.; Miret-Art\'es, S. Stochastic Bohmian and Scaled Trajectories. \emph{Found. Phys.}  \textbf{2022}, \emph{52}, 78.
%
\bibitem{Feng}
Xiao-Feng, P.; Yuan-Ping, F. {\it Quantum Mechanics in Nonlinear Systems}; World Scientific: Hong Kong, China, 2005.
%
\bibitem{Schi-PR-1962}
Schiller, R. Quasi-classical theory of the nonspinning electron. \emph{Phys. Rev.} \textbf{1962}, \emph{125}, 1100.
%
\bibitem{MoMi-JPA-2023}
Mousavi, S.V.;  Miret-Art\'es, S. Different routes to the classical limit of backflow. \emph{J. Phys. A Math. Theor.}  \textbf{2023}, \emph{55}, 475302.
%
\bibitem{SaBo-EPJD-2007}
Sanz, A.S.; Borondo, F. A quantum trajectory description of decoherence. \emph{Eur. Phys. J. D}  \textbf{2007}, \emph{44}, 319.
%
\bibitem{LuSa-AP-2015}
Luis, A.; Sanz, A.S. What dynamics can be expected for mixed states in two-slit experiments? \emph{Ann. Phys.}  \textbf{2015}, \emph{357}, 95.
%
\bibitem{MaBi-JCP-2001}
Maddoxa, J.B.; Bittner, E.R. Quantum relaxation dynamics using Bohmian trajectories. I. General formulation. \emph{J. Chem. Phys.} \textbf{2001}, \emph{115},  6309.
%
\bibitem{BuCe-JCP-2001}
Burghard, I.; Cederbaum, L.S. Hydrodynamic equations for mixed quantum states. I. General formulation. \emph{J. Chem. Phys.} \textbf{2001}, \emph{115}, 10303 .
%
\bibitem{Mo-PCPS-1949}
Moyal, J.E. Quantum mechanics as a statistical theory. P\emph{roc. Cam. Phil. Soc.}  \textbf{1949}, \emph{45}, 99--123.
%
\bibitem{Hi-arXiv-2014}
Hiley, B.J. Moyal's Characteristic Function, the Density Matrix and von Neumann's Idempotent. \emph{arXiv} \textbf{2014},  arXiv:1408.5680.%MDPI: please add year.
%
\bibitem{Hi-JCE-2015}
Hiley, B.J. On the relationship between the Wigner–Moyal approach and the quantum operator algebra of von Neumann. \emph{J. Comput. Electron}  \textbf{2015}, \emph{14}, 869--878. 
%
\bibitem{Sakurai-book-1994}
Sakurai, J.J. {\it Modern Quantum Mechanics Revised Edition}; 
Tuan, S.F., Ed.; Addison-Wesley Publishing Company: Boston, MA, USA, 1994.
%
%\bibitem{RiScMaVaBa-PRA-2014}
%C. D. Richardson, P. Schlagheck, J. Martin, N. Vandewalle and T. Bastin,  Phys. Rev. A~ \emph{89}, 032118 (2014).
%
\bibitem{Ro-AJP-1964}
Rosen, N. The Relation Between Classical and Quantum Mechanics. \emph{Am. J. Phys. } \textbf{1964}, \emph{32}, 597. 

\bibitem{Wigner}
Wigner, E. On the Quantum Correction For Thermodynamic Equilibrium. \emph{Phys. Rev.} \textbf{1932}, \emph{40}, 749.
%
\bibitem{GrSt-book-1998} 
Grosche, C.; Steiner, F. {\it Handbook of Feynman Path Integrals}; Springer: Berlin/Heidelberg, Germany, 1998.
%
\bibitem{Salva2008} Sanz, A.S.; Miret-Art\'es, S. A trajectory-based understanding of quantum interference. \emph{J. Phys. A Math. Theor.}  \textbf{2008}, \emph{41}, {435303}.%MDPI: We have corrected the page information, please confirm.
%
\bibitem{Le-book-2008}
Leavens, C.R. Bohm Trajectory Approach to Timing Electrons. In {\it Time in Quantum Mechanics}; Muga, J., Mayato, R.S., Egusquiza, I., Eds.; Lecture Notes in Physics; Springer: Berlin/Heidelberg, Germany, 2008;  Volume 734.
%
\bibitem{Zwanzig-2001} 
Zwanzig, R. {\it Nonequilibrium Statistical Mechanics}; Oxford University Press: Oxford, UK, 2001.
%
\bibitem{Ni-FPL-2006}
Nikolic, H. Classical Mechanics Without Determinism. \emph{Found. Phys. Lett. } \textbf{2006}, \emph{19}, 553. 
%




\end{thebibliography}
\end{document}